\documentclass[preprint]{emulateapj}

\usepackage{amsmath}
\usepackage{graphicx}
\usepackage{epstopdf}
\usepackage{hyperref}

\newcommand\beq{\begin{equation}}
\newcommand\eeq{\end{equation}}
\newcommand\beqar{\begin{eqnarray}}
\newcommand\eeqar{\end{eqnarray}}

\defcitealias{pv08}{PV08}
 
\shorttitle{Blazar Spectral Breaks and the EGB} 
\shortauthors{Venters \& Pavlidou}
 
\begin{document}
\DeclareGraphicsExtensions{.pdf,.gif,.jpg}

\title{The Effect of Blazar Spectral Breaks on the Blazar Contribution to the Extragalactic Gamma-ray Background}

\author{Tonia M. Venters\altaffilmark{1,3} and Vasiliki Pavlidou\altaffilmark{2,4}}
\altaffiltext{1}{Astrophysics Science Division, NASA Goddard Space Flight Center, Greenbelt, MD 20771; \href{mailto:tonia.m.venters@nasa.gov}{tonia.m.venters@nasa.gov}}
\altaffiltext{2}{Department of Astronomy, The California Institute of Technology, Pasadena, CA 91125}
\altaffiltext{3}{NASA Postdoctoral Program Fellow}
\altaffiltext{4}{Einstein Fellow}

\begin{abstract}
The spectral shapes of the contributions of different classes of unresolved gamma-ray emitters can provide insight into their relative contributions to the extragalactic gamma-ray background (EGB) and the natures of their spectra at GeV energies. We calculate the spectral shapes of the contributions to the EGB arising from BL Lacertae type objects (BL Lacs) and flat-spectrum radio quasars (FSRQs) assuming blazar spectra can be described as broken power laws. We fit the resulting total blazar spectral shape to the \emph{Fermi} Large Area Telescope measurements of the EGB, finding that the best-fit shape reproduces well the shape of the \emph{Fermi} EGB for various break scenarios. We conclude that a scenario in which the contribution of blazars is dominant cannot be excluded on spectral grounds alone, even if spectral breaks are shown to be common among \emph{Fermi} blazars. We also find that while the observation of a featureless (within uncertainties) power-law EGB spectrum by \emph{Fermi} does not necessarily imply a single class of contributing unresolved sources with featureless individual spectra, such an observation and the collective spectra of the separate contributing populations determine the ratios of their contributions. As such, a comparison with studies including blazar gamma-ray luminosity functions could have profound implications for the blazar contribution to the EGB, blazar evolution, and blazar gamma-ray spectra and emission.
 \end{abstract}

\maketitle 

\section{Introduction}\label{sec:intro}

The gamma-ray sky as currently observed by the Large Area Telescope (LAT) on board the \emph{Fermi Gamma-Ray Space Telescope} consists of resolved point sources (such as active galactic nuclei (AGNs), pulsars, and star-forming galaxies), transient gamma-ray sources (e.g., gamma-ray bursts), and the diffuse gamma-ray radiation comprised of emission from the Galaxy and the extragalactic gamma-ray background (EGB). The origins of the EGB are, as yet, unknown; however, it is expected that unresolved, extragalactic point sources provide a sizable contribution to the EGB.

In its first year of taking data, \emph{Fermi} observed $1451$ resolved point sources with significance greater than $5\sigma$ (First \emph{Fermi}-LAT catalog (1FGL); \citealt{lat10b}), of which $573$ were associated with blazars (AGNs for which the jet is closely aligned with the observer's line of sight \citep{bla79}). Thus, blazars constitute the largest class of astrophysical objects associated with gamma-ray sources, and it has long been suspected that unresolved blazars should provide a substantial contribution to the EGB \citep{pad93,ste93,sal94,chi95,ss96a,kaz97,chi98,sre98,muk99,muc00,gio06,nt06,der07,kne08,ino08,ino10,ino09,sclat10,ste10}. 

It should then, perhaps, come as no surprise that the \emph{Fermi}-LAT, with its improved sensitivity enabling it to observe many more blazars, measured the integrated intensity of the EGB to be substantially lower than that measured by the Energetic Gamma-Ray Experiment Telescope (EGRET) aboard the \emph{Compton Gamma-ray Observatory} in the 1990s ($1.03\ (\pm \ 0.17)\times 10^{-5} \mbox{ cm}^{-2} \mbox{ s}^{-1} \mbox{ sr}^{-1}$ for \emph{Fermi} versus $1.45\ (\pm \ 0.05)\times 10^{-5} \mbox{ cm}^{-2} \mbox{ s}^{-1} \mbox{ sr}^{-1}$ for EGRET; \citealt{sre98,lat10a}).  However, exactly how much of the change in the EGB integrated intensity is due to the ability to resolve many more blazars remains unclear. On the one hand, since the \citet{sre98} determination of the EGRET EGB, models of the galactic foreground emission have been updated to reflect recent observations of the interstellar medium \citep{lat10a}, accounting for at least some of the change. On the other hand, the determination of the distribution of blazars with respect to luminosity and redshift (the blazar gamma-ray luminosity function, GLF) and by extension, their contribution to the EGB directly from their \emph{observed} flux distribution (such as that performed in \citealt{sclat10}) is non-trivial, for four reasons.  

First, blazars are variable at gamma-ray energies and without knowledge of the degree to which the gamma-ray luminosity of a blazar changes during the flaring state and the blazar duty cycle, the reconciliation of the \emph{observed} blazar source counts with model GLFs is impossible \citep{ss96a}. This is because the increase in flux of a blazar (sometimes up to an order of magnitude; see e.g., \citealt{mcl96,muk97,nol03,ver04}) during the flaring period introduces a selection effect in that flaring blazars are easier to detect than quiescent blazars, and quiescent blazars would have to be relatively bright and nearby to be detectable. The largest impact on the \emph{observed} blazar source counts will be at fluxes just above the \emph{Fermi}-LAT sensitivity as some blazars below the threshold make it into the sample because they are flaring. While the magnitude of the impact on the \emph{observed} blazar source counts depends on the blazar GLF and the blazar duty cycle, the effect will be to \emph{flatten} the faint-end slope of the observed blazar source counts (as blazars move from lower flux bins to higher ones) leading to an \emph{underestimation} of the blazar contribution to the EGB as extrapolated from the observed source counts alone. \citet{sclat10} concluded that since the peak-to-mean flux ratio for most \emph{Fermi} sources is $\sim$ a factor of two, that there are no large systematic uncertainties due to blazar variability. However, a small peak-to-mean ratio is expected for blazars that are observed mostly in a flaring state, and as most of the observed blazars are fainter blazars that will be more represented by flaring blazars, then while it will be true that most blazars will have a small peak-to-mean ratio, it is not necessarily the case that the effect of blazar variability will be small.

Second, the large angular resolution of the \emph{Fermi}-LAT at lower energies ($\sim 5^{\circ}$ at $100$ MeV) could introduce significant source confusion for GLF models that predict high blazar densities resulting in fewer resolved blazars and a higher contribution to the EGB at lower energies \citep{ss99b,ste10}. The impact of source confusion on observed source counts depends on the source density as given by the blazar GLF. As such, accounting for source confusion requires \emph{a priori} knowledge of the source density, which is exactly what the observer wishes to determine. Thus, source confusion further complicates analyses of the blazar contribution to the EGB based solely on observed source counts \citep[for a detailed discussion of source confusion and the difference between the definition employed in this discussion and that employed by \citealt{sclat10}, see][]{ste10}. 

Third, as demonstrated in \citet{sclat10}, the reconstructed fluxes of the individual sources are subject to a considerable amount of uncertainty, particularly at the faint end (see their Figure 6).  Finally, the model of the \emph{Fermi}-LAT detection efficiency employed by \citet{sclat10} assumed power-law spectra for the sources\footnote{It should also be noted that the model for \emph{Fermi}-LAT detection efficiency (particularly at energies below $\sim 300$ MeV where multiple scattering in the detector becomes important) is subject to change during the course of observations by \emph{Fermi}-LAT.}. As such, these uncertainties complicate the determination of the distribution of blazars with respect to luminosity and redshift (the GLF) and the determination of the blazar contribution to the EGB from their \emph{observed} flux distribution. Notably, \citet{ste10} found that contrary to the conclusion of \citet{sclat10}, the observed flux distribution of \emph{Fermi} blazars \emph{does not}, as yet, rule out the possibility that the EGB is dominated by emission from unresolved blazars \citep[see also][]{aba10}. Thus, the contribution of still unresolved blazars to the EGB remains in dispute.

Additional information about the contributions to the EGB can be obtained through studying the \emph{shape} of its energy spectrum, and the shapes of the collective intensity spectra of suspected contributors (\citealp{pv08}, hereafter PV08; \citealp{ven09,ven10}).
Analysis of the first year of \emph{Fermi} data yielded an EGB spectrum consistent with a featureless power law with spectral index $\Gamma \sim 2.4$ \citep{lat10a}. Upon first reflection, the \emph{Fermi} EGB spectrum appears to be consistent with the hypothesis that the EGB is dominated by emission from unresolved blazars since the mean spectral index for blazars, $\Gamma_0$, is also $\sim 2.4$.  However, several effects complicate this simple picture. 

First, the astrophysical population of blazars is actually composed of two separate sub-populations (flat-spectrum radio quasars, FSRQs, and BL Lacs) with distinct spectral properties ($\Gamma_0 \sim 2.45$ for FSRQs and $\Gamma_0 \sim 2.2$ for BL Lacs; \citealt{sclat10}).  Second, even within a given sub-population of blazars, the spectral indices of individual blazars form a distribution with some spread \citep{ss96a,vp07}, which causes the collective spectrum of unresolved blazars to curve as harder blazar spectra become more important at higher energies \citep{ss96a,pv08}. On the other hand, this effect is somewhat mitigated by the spectral bias introduced by the fact that blazars with harder spectral indices are easier to observe in a flux-limited survey\footnote{It should be noted that the \emph{Fermi} survey is not exactly a flux-limited survey due to the non-uniformity of the total diffuse background throughout the sky.  As such, spectral bias is more significant in the \emph{Fermi} survey than typical for a truly flux-limited survey.  However, as evidenced from the analysis of the mean spectral indices performed by \citet{sclat10}, this appears to have more of an effect on BL Lacs than FSRQs.}, and thus, are more likely to be resolvable and not play as big a role in producing the EGB.  Third, just as unresolved blazars are expected to contribute to the EGB, unresolved members of other known astrophysical gamma-ray emitters (such as star-forming and starburst galaxies) should also contribute, but with spectra that are substantially different from those of blazars \citep{pf02,fie10,lacki10,mak10,ste10} and may not even resemble power laws at gamma-ray energies.  Finally, recent observations conducted by {\it Fermi} indicate that even blazar spectra can break at $\sim$ GeV energies and would no longer be describable by simple power laws \citep{lat09a,ino08,ino10,ino09}.  

The intuitive conclusion would be that the combination of these effects should cause the energy spectrum of the EGB intensity to exhibit features \citep{pf02}.  The observation instead of an  EGB with a single, featureless power-law spectrum begs the question: {\em is the lack of observed spectral features necessarily indicative of a single dominant source population with individual unbroken power-law spectra in the energy range of $\geq 300$ MeV?} If not, then {\em what can such a featureless power-law EGB spectrum tell us about the relative contributions of the separate populations and how do they depend on their collective spectra?} The answer to these questions provide novel constraints on blazar GLFs and thus could have profound implications for the cosmological properties of known astrophysical gamma-ray emitters, their corresponding contributions to the EGB, and the general properties of their spectra at gamma-ray energies. 

In this paper, we calculate the collective spectrum of unresolved blazars with individual spectra exhibiting broken power laws.  In so doing, we seek to determine whether such a population of unresolved blazars can result in a featureless power-law spectrum resembling that of the EGB. If so, we could then investigate the implications for the relative contributions of FSRQs and BL Lacs, which could, in turn, have implications for the cosmological properties of blazars. In Section 2, we present the formalism of the calculation of the spectral shape of the collective unresolved blazar emission. In Section 3, we discuss the inputs of the calculation and their uncertainties. In Section 4, we present the results of the calculation, and we discuss them in Section 5.

\section{Formalism}\label{sec:formalism}

To calculate the collective spectrum of unresolved blazars, we follow the formalism as outlined in \citetalias{pv08} with one major difference.  Instead of taking blazar spectra to be simple power laws over gamma-ray energies ($F \propto E^{-\Gamma}$, where $\Gamma$ is the \emph{photon} spectral index at gamma-ray energies), blazar spectra are taken to be smoothly broken power laws:
\begin{equation}
F_E(E) = F_0\left[\left(\frac{E}{E_b}\right)^{\Gamma_1n}+\left(\frac{E}{E_b}\right)^{\Gamma_2n}\right]^{-1/n},
\end{equation}
where $F_E(E)$ is the differential photon flux in units of photons per unit area per unit energy per unit time, $E_b$ is the break energy, $\Gamma_1$ is the low-energy slope, $\Gamma_2$ is the high-energy slope, and $n$ quantifies the sharpness of the transition from the low-energy power law to the high-energy power law. For the purposes of this paper, we take $n=1$. The total flux, $F$, of photons with energies greater than some fiducial energy,  $E_f$, is found by integrating $F_E(E)$ over energy,
\begin{equation}
F = F_0\int_{E_f}^{\infty}\left[\left(\frac{E}{E_b}\right)^{\Gamma_1n}+\left(\frac{E}{E_b}\right)^{\Gamma_2n}\right]^{-1/n}dE\,.
\end{equation}
 For the purposes of this paper, we take $E_f = 100$ MeV. Then, the contribution of a single unresolved blazar to the EGB is
\begin{equation}
I_1 = \frac{F\left[(E/E_b)^{\Gamma_1n}+(E/E_b)^{\Gamma_2n}\right]^{-1/n}}{4\pi\int_{E_f}^{\infty}\left[(E/E_b)^{\Gamma_1n}+(E/E_b)^{\Gamma_2n}\right]^{-1/n}dE} 
\end{equation}
where the flux of one source is uniformly distributed over the entire sky in anticipation of an isotropically distributed cosmological population and $I$ has units of photons per unit area per unit energy per unit time per unit solid angle.

Following \citetalias{pv08}, we characterize the flux distribution of unresolved blazars as a function $g(F)$ and the distribution of blazar spectral indices (or spectral index distribution, SID) as a function $p(\Gamma)$, where $\Gamma_1 = \Gamma - \Delta\Gamma_1$, $\Gamma_2 = \Gamma + \Delta\Gamma_2$, and $p(\Gamma)$, $\Delta\Gamma_1$, and $\Delta\Gamma_2$ are determined from observations (see Section \ref{sec:inputs}).  Then, the total contribution of unresolved blazars to the EGB is given by
\begin{equation}\label{eqn:EGB}
I(E) = \int_{-\infty}^{\infty}d\Gamma\int_{0}^{F_{\rm min}}dFg(F)I_1p(\Gamma)\,,
\end{equation}
where $F_{\rm min}$ is the sensitivity of the gamma-ray telescope under consideration. For the first year of {\it Fermi} data, $F_{\rm min} \sim 2 \times 10^{-9} \mbox{ photons cm}^{-2} \mbox{ s}^{-1}$.

Equation (\ref{eqn:EGB}) can be characterized in terms of factors that determine the overall magnitude of the unresolved blazar contribution to the EGB (flux terms) and factors that determine the overall shape of the blazar contribution (spectral index terms). For a carefully chosen definition of $p(\Gamma)$ (determined from the analysis of spectral indices of the {\em flux-limited} sample of {\it Fermi} blazars that accounts for the spectral bias inherent in a flux-limited catalog; see Section \ref{sec:inputs} and \citealp{ven09}), the magnitude and shape terms decouple, and Equation (\ref{eqn:EGB}) can be rewritten as
\begin{multline}\label{eqn:EGB-decoupled}
I(E) = \\ 
I_0 \!\! \int_{-\infty}^{\infty} \!\!\!\!\!\!\! d\Gamma \, p(\Gamma)\frac{\left[(E/E_b)^{(\Gamma - \Delta\Gamma_1)n}+(E/E_b)^{(\Gamma + \Delta\Gamma_2)n}\right]^{-1/n}}{\mathcal{S}(E_f,\Gamma)}\,,
\end{multline}
where $I_0$ is a normalization constant depending on the flux distribution of unresolved blazars and
\begin{multline}
\mathcal{S}(E_f,\Gamma) = \\
\int_{E_f}^{\infty} \!\!\! dE \left[\left(\frac{E}{E_b}\right)^{(\Gamma - \Delta\Gamma_1)n}+\left(\frac{E}{E_b}\right)^{(\Gamma + \Delta\Gamma_2)n}\right]^{-1/n}.
\end{multline}

\section{Inputs}\label{sec:inputs}

As demonstrated in \citetalias{pv08}, the unresolved blazar contribution to the EGB {\em is not just a question of magnitude, but also of spectral shape}, and the spectral shape is sensitive to the distribution of blazar spectral indices at GeV energies. Both \citetalias{pv08} and \citet{sclat10} assumed that blazar spectra at gamma-ray energies are power laws and did not account for possible breaks in the spectra.  In this paper,  blazar spectra take the forms of smoothly-broken power laws (see Section \ref{sec:formalism}).  

\begin{figure}[t]
\begin{center}
\resizebox{3.0in}{!}
{\includegraphics[trim = 0 0 0 0.5mm, clip]{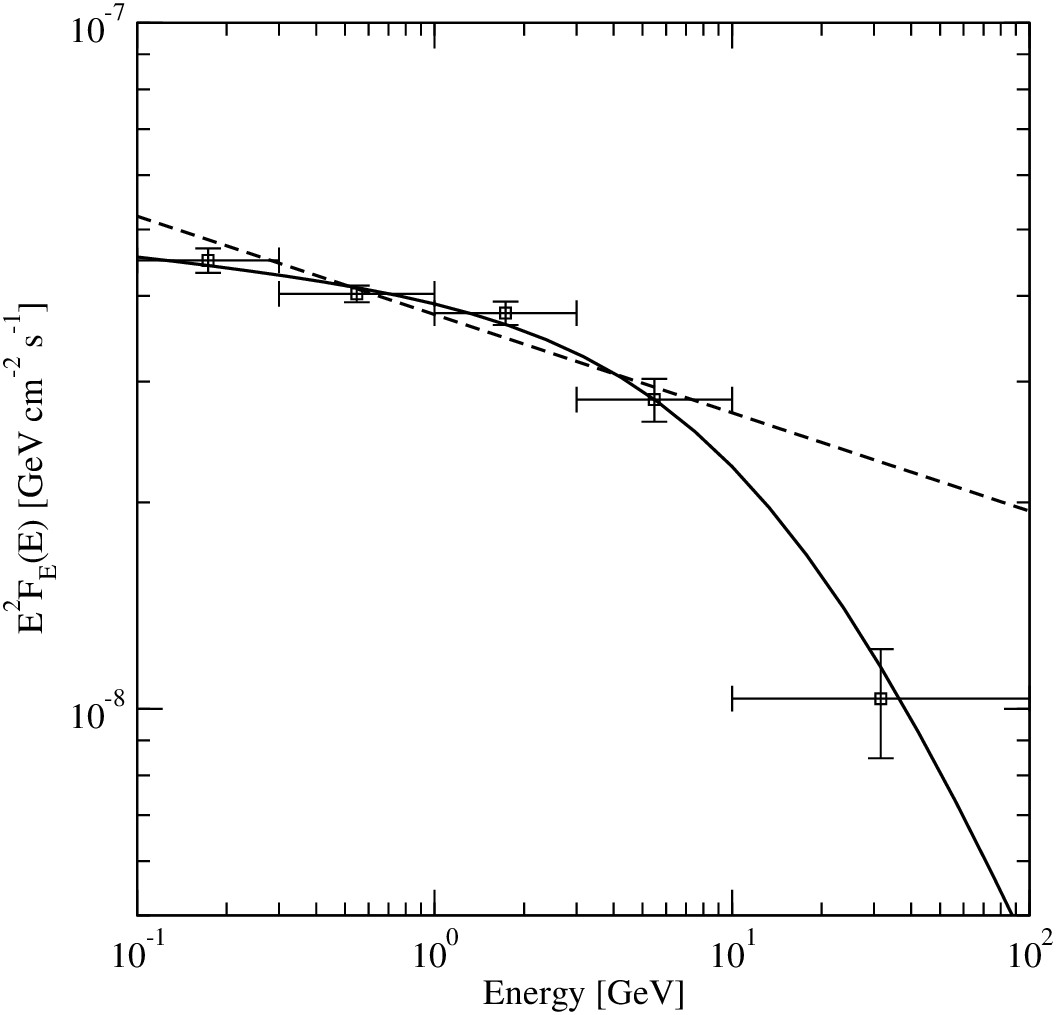}}
\resizebox{3.0in}{!}
{\includegraphics[trim = 0 0 0 0.5mm, clip]{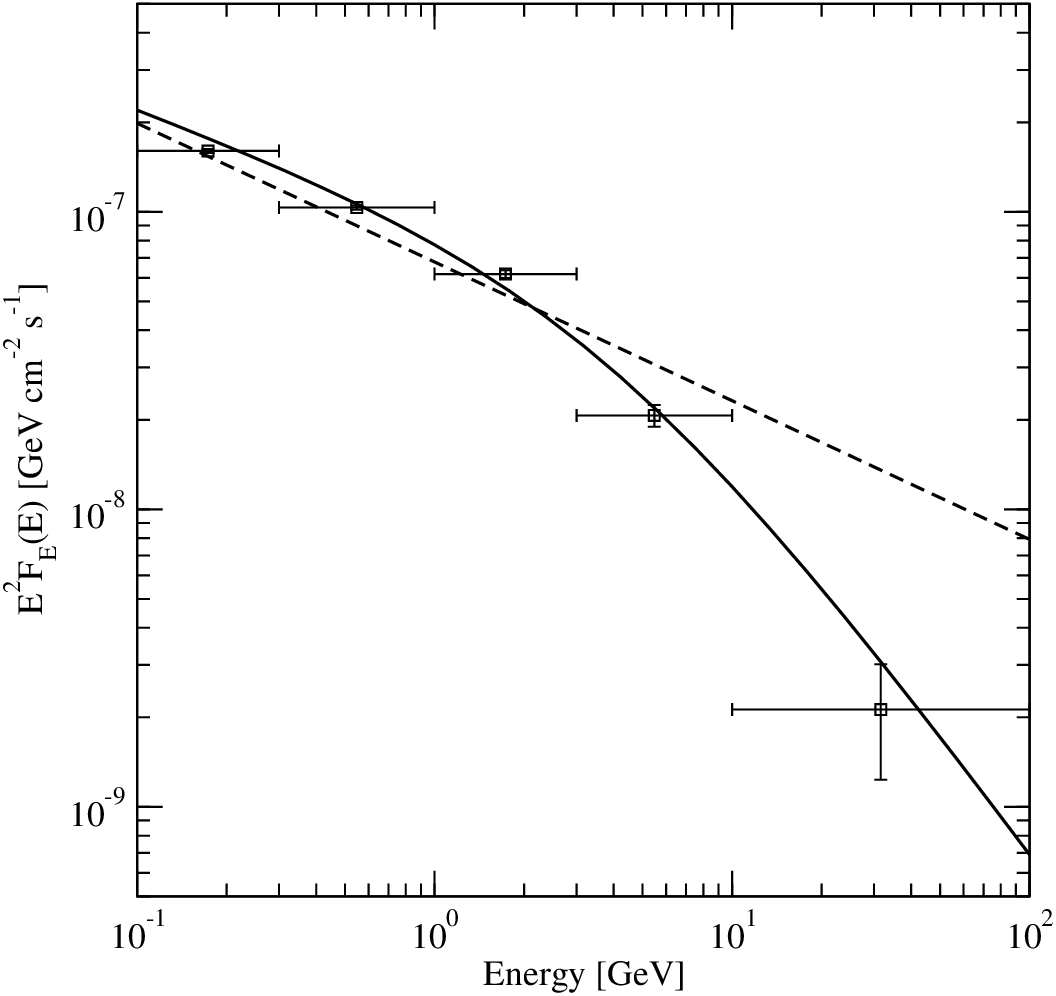}}
\caption{Sample broken power-law spectra for blazars. Top: PKS0235+164 measured spectrum with broken power-law (solid) and single power-law spectra (dashed; $\Gamma = 2.1433$). Bottom: 3C454.3 measured spectrum with broken power-law and single power-law spectra ($\Gamma = 2.4662$).\label{fig:blazarspectra}}
\end{center}
\end{figure}

For each sub-population of blazars, we determine the SID, $p(\Gamma)$, from the likelihood analysis of \citet{vp07} fitting blazars from the {\it Fermi}-LAT First Catalog of AGNs \citep{agnlat10} to Gaussian SIDs accounting for errors in measurement of individual spectral indices.  Due to the survey bias towards harder spectral indices present in {\it Fermi} data, we applied the likelihood analysis only to the subset of blazars with photon fluxes $\gtrsim 7 \times 10^{-8}\mbox{ photons cm}^{-2} \mbox{ s}^{-1}$ and galactic latitudes $\gtrsim 10^{\circ}$ (as per \citealt{sclat10}). We determined that the maximum-likelihood Gaussian SID can be characterized by a mean ($\Gamma_0$) and a spread ($\sigma_0$) with maximum-likelihood parameters determined to be $\Gamma_0 = 2.45$ and $\sigma_0 = 0.16$ for FSRQs and $\Gamma_0 = 2.17$ and $\sigma_0 = 0.23$ for BL Lacs, which are similar to the findings of \citet{sclat10}.  Based on 1FGL spectra of two prominent blazars from the \emph{Fermi}-LAT Bright AGN Sample \citep[][; see Figure \ref{fig:blazarspectra}]{lat09a}, we model the spectral breaks by taking $\Delta\Gamma_1 = 0.1$ and $\Delta\Gamma_2 = 0.9$.  As in the measured blazar spectra, we treat the break energies distinctly for FSRQs and BL Lacs.  We took the break energy, $E_b$, to be $4$ GeV for FSRQs and $15$ GeV for BL Lacs.  In so doing, we consider two cases:  
\begin{itemize}
\item[] {\it Scenario 1}. - Blazars within a population evolve such that their break energies are \emph{observed} to be roughly the same.
\item[] {\it Scenario 2}. - The break energies of the \emph{intrinsic} spectra of blazars within a subpopulation are the same, but because of redshift effects, they are \emph{observed} to be different.
\end{itemize}
In order to perform the calculation in the Scenario 2, we separately model the contribution arising from different redshift bins from an assumed GLF model for a given subpopulation of blazars and then determine the composite spectrum in the final step. For the purposes of this analysis, we assume the best-fit pure luminosity evolution model of \citet{nt06} for FSRQs and their best-fit luminosity-dependent density evolution model for BL Lacs\footnote{In so doing, we account for the observation that BL Lacs are on average situated at lower redshifts than FSRQs.}. In this manner, we seek to investigate the impact of changing the GLF model(s) of the blazars. We should also note that alternative statistical treatments on more extensive data sets as in \citet{spclat10} could result in different estimates of the break energies and changes in spectral index. 

Another alternative treatment performed by \citet{sclat10} consists of stacking the spectra of observed blazars. However, the analysis was performed on the \emph{flux-limited} sample of {\it Fermi} blazars, which consists almost entirely of FSRQs (hence, the similarity between the stacked spectrum for the flux-limited sample of blazars and that of FSRQs). Moreover, while such an analysis is effective in determining the \emph{average} spectral index of a given population of blazars, it reveals nothing about the spectral properties of the \emph{population} of blazars (e.g., the spread in the SID, spectral breaks). Furthermore, as the {\it Fermi}-LAT survey is biased against sources with softer spectra, it is certainly biased against sources exhibiting significant breaks in their spectra. We also note that despite finding that the SIDs of blazars have non-zero spreads, \citet{sclat10} did not account for these spreads in their calculated spectra of the blazar contributions to the EGB as evidenced by the fact that these spectra, as well as their estimated uncertainties, are still power laws (see Figures 18, 19, and 20 of \citealt{sclat10}). Similarly, \citet{sclat10} did not account for spectral breaks in their calculated spectra of the blazar contributions to the EGB. Inclusion of these effects would introduce curvature in the spectra of the blazar contributions to the EGB and not result in the power laws indicated by \citet{sclat10}.

As in PV08, we do not include information about the magnitude of the unresolved blazar contribution to the EGB.  Instead, we normalize the collective spectrum of each sub-population in order to best fit (as determined using a $\chi^2$-analysis) the measured spectrum of the EGB from the {\it Fermi}-LAT first year data \citep{lat10a}.  For the composite spectrum of FSRQs and BL Lacs, we normalize the collective spectrum of each sub-population such that their \emph{total} composite spectrum is the best-fit spectrum to the measured spectrum.

We should note that while applying the aforementioned cuts to the sample accounts for much (though likely not all; see \citealt{ven09}) of the survey spectral bias, doing so also significantly reduces the sample size and will ultimately introduce more uncertainty in the determination of the likelihood parameters. Furthermore, while in the \emph{flux-limited} sample of blazars, spectral index and flux do appear to decouple \citep[see Figure 1 of][]{sclat10}, we expect it to be only approximately correct. A more detailed calculation, however, requires a statistical analysis to determine the \emph{intrinsic} blazar SID, which would, in turn, require knowledge of the blazar GLFs since the survey spectral bias depends on the redshift and luminosity distributions of blazars\footnote{For this same reason, we also neglect source confusion at low energies arising from the increase in \emph{Fermi}-LAT point-spread function below $1$ GeV.} \citep{ven09}. In the case of FSRQs, applying the high-flux cut changes the SID very little \citep{sclat10,ste10}, so we do not expect the collective spectrum of unresolved FSRQs to be appreciably different from that determined from the resolved FSRQs. On the other hand, in the case of BL Lacs, the sparseness of the population at fluxes above the cut introduces a considerable margin of error. As such, the collective spectrum of unresolved BL Lacs could be harder or softer than that expected from the resolved BL Lacs. As BL Lacs are, on average, harder than FSRQs, they will likely contribute most significantly at the higher energies, and hence, the differences in spectra between resolved and unresolved BL Lacs will mostly impact the collective unresolved blazar spectrum at higher energies. 

Finally, we should note that recent multi-wavelength observations conducted by {\it Fermi} and other telescopes suggest that blazars can be further subdivided into the categories of low synchrotron peak, intermediate synchrotron peak, and high synchrotron peak \citep{sedlat10}.  In principle, one could gain further insight by applying this procedure to each subdivision as there is some indication that these categories are also spectrally distinct at gamma-ray energies \citep{sedlat10}. Notably, as demonstrated in \citet{spclat10}, most high-synchrotron peaked BL Lacs (HBLs) do not exhibit spectral breaks in \emph{Fermi}-LAT energy range while those HBL breaks that have been observed are different in character (going from soft to hard rather than hard to soft). However, in light of the cuts on galactic latitude and flux, further subdividing the sample of blazars could lead to small sample sizes in some of the categories, and consequently to poor SID parameter determination.  Thus, for the purposes of this analysis, we retain the original classification scheme of FSRQs and BL Lacs.  In any case, given that BL Lacs likely contribute most significantly at higher energies, the effect of the HBL spectra is likely to improve the fit at high energies, though a more detailed study is in order as more data become available. One might also be concerned about systematic changes in blazar spectral indices while flaring; however, analyses of EGRET blazar spectral indices found no evidence of such systematic changes in spectral index with flaring \citep{nan07,vp07}, and {\it Fermi} observations of individual blazars have thus far revealed no systematic changes in spectral index with time or flux \citep{09pks1454-354,10pks1510-089,103C454.3}.

\section{Results}\label{sec:results}

\begin{figure}[t]
\begin{center}
\resizebox{3.0in}{!}
{\includegraphics[trim = 0 0 0 0.5mm, clip]{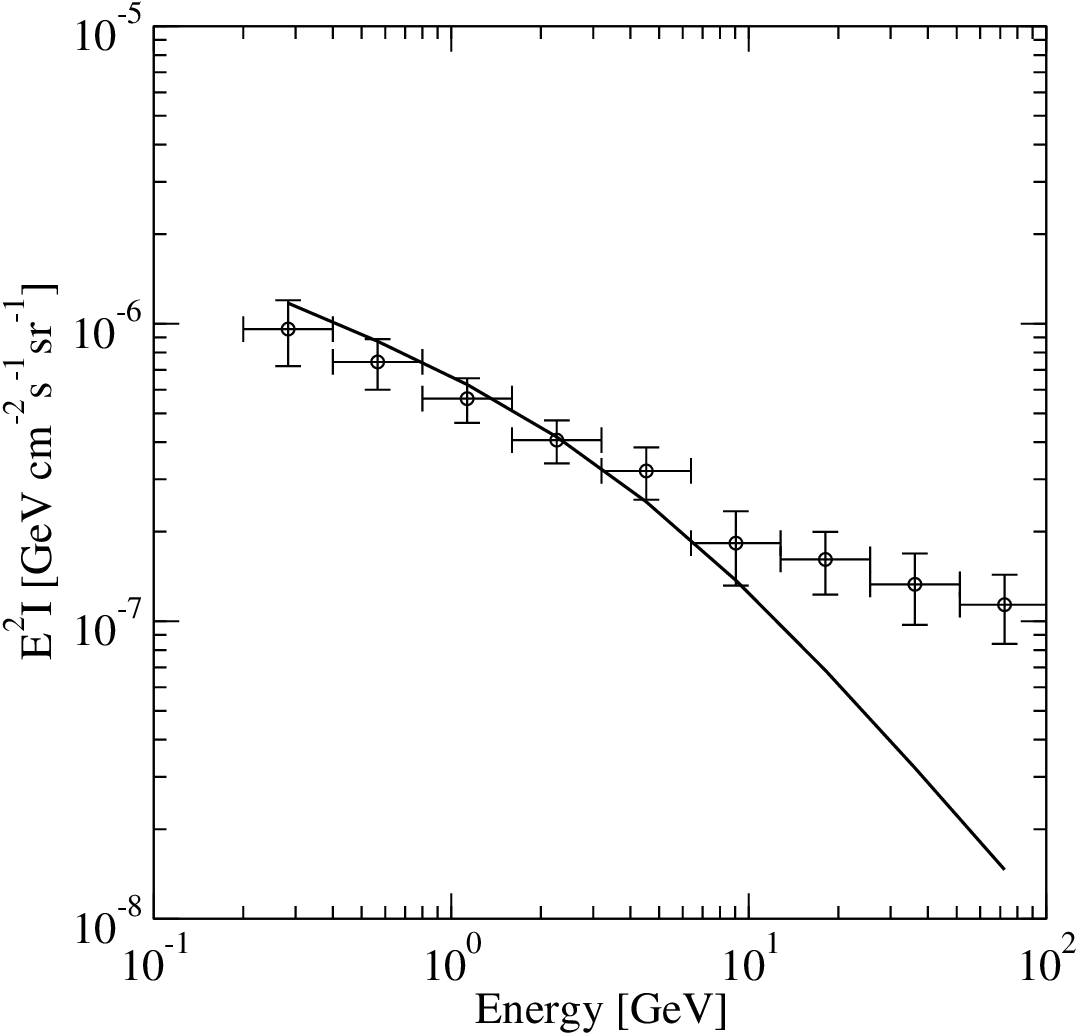}}
\resizebox{3.0in}{!}
{\includegraphics[trim = 0 0 0 0.5mm, clip]{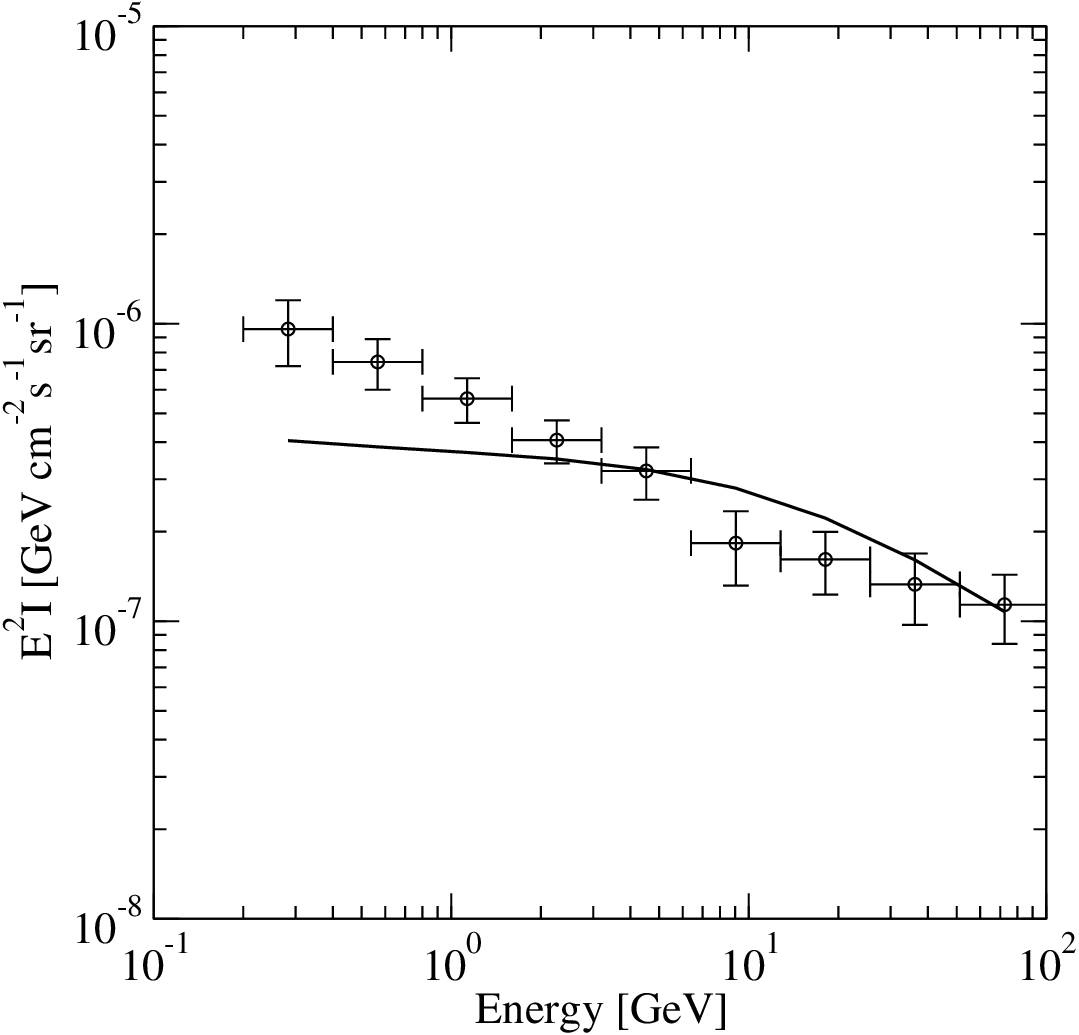}}
\caption{Best-fit shapes of the individual collective intensities of the sub-populations of unresolved blazars. Top: the best-fit shape for FSRQs. Bottom: the best-fit shape for BL Lacs.  Data points: spectrum of the EGB as measured by the {\it Fermi}-LAT \citep{lat10a}. \label{fig:blazarEGB}}
\end{center}
\end{figure}

\begin{figure}[t]
\begin{center}
\resizebox{3.0in}{!}
{\includegraphics[trim = 0 0 0 0.5mm, clip]{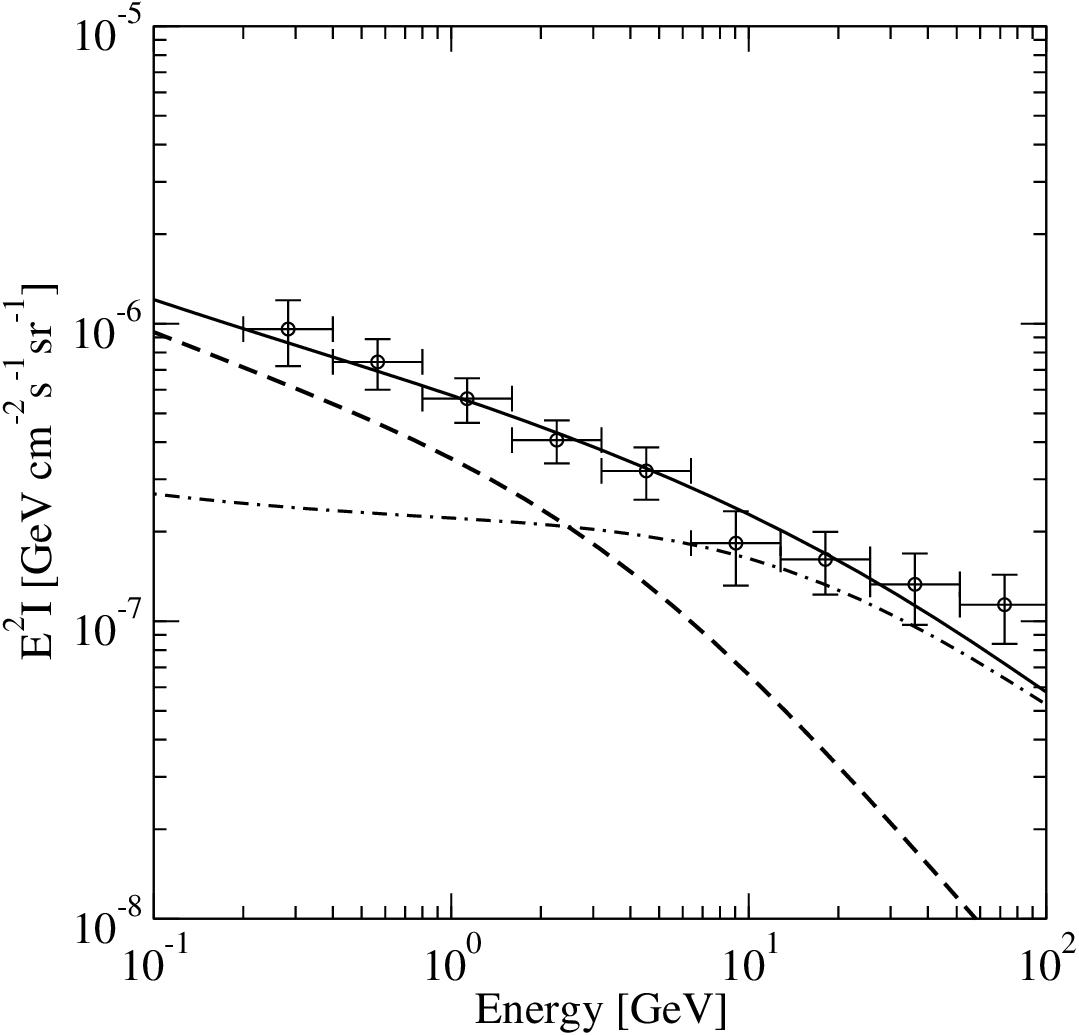}}
\caption{Best-fit shape of the total collective intensity of unresolved blazars in Scenario 1 as described in Section \ref{sec:inputs}. Dashed line: the contribution from FSRQs. Dot-dashed line: the contribution from BL Lacs. Solid line: the shape of the combined population of FSRQs and BL Lacs. Data points: same as in Figure \ref{fig:blazarEGB}\label{fig:bestcombinedEGB}}.
\end{center}
\end{figure}

\begin{figure}[t]
\begin{center}
\resizebox{3.0in}{!}
{\includegraphics[trim = 0 0 0 0.5mm, clip]{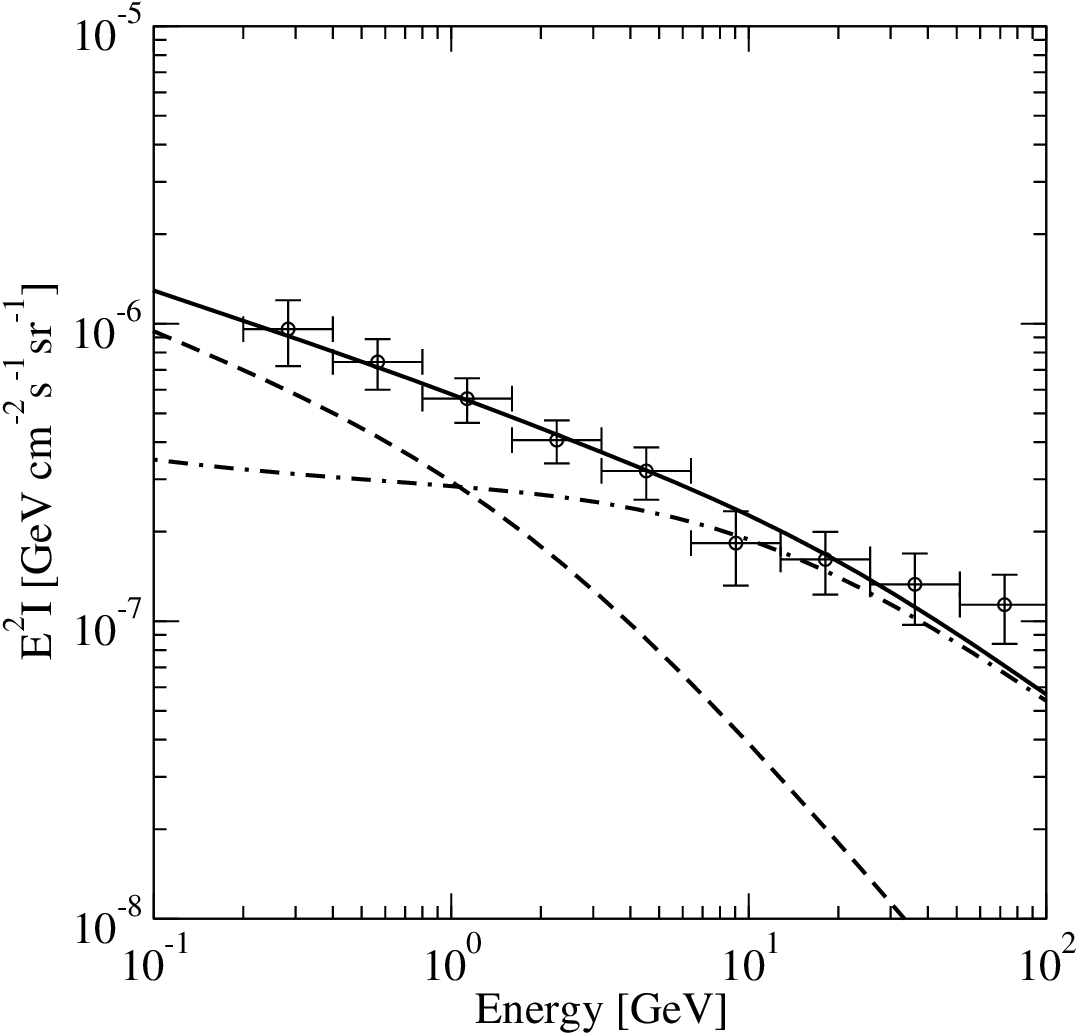}}
\caption{Same as in Figure \ref{fig:bestcombinedEGB}, but for Scenario 2 as described in Section \ref{sec:inputs}.\label{fig:zdepcombinedEGB}}
\end{center}
\end{figure}

\begin{figure}[t]
\begin{center}
\resizebox{3.0in}{!}
{\includegraphics[trim = 0 0 0 0.5mm, clip]{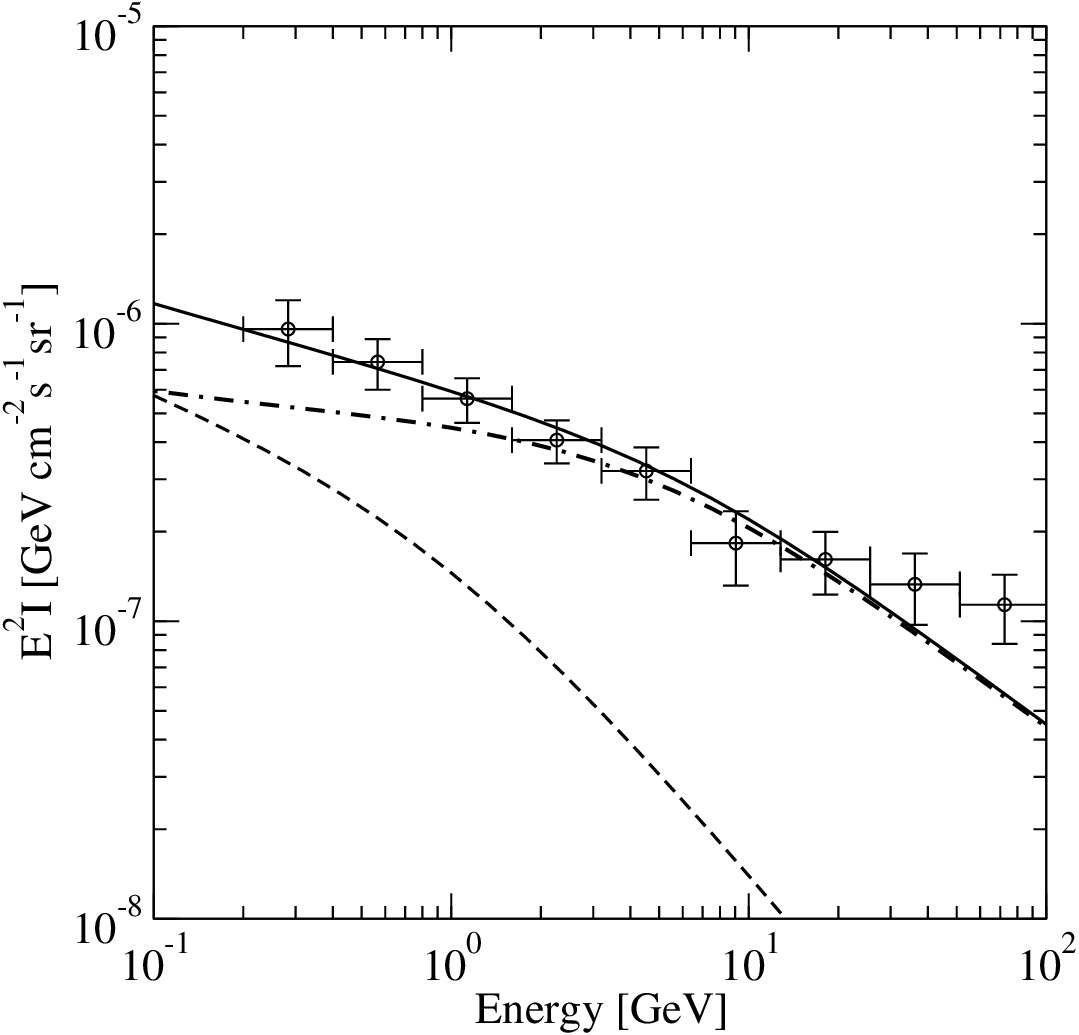}}
\caption{Same as in Figure \ref{fig:bestcombinedEGB}, but for lower break energies.\label{fig:altcombinedEGB}}
\end{center}
\end{figure}

The spectral shapes of the contributions of unresolved blazars to the EGB as determined from the Gaussian SIDs and broken power-law gamma-ray spectra discussed in Section \ref{sec:inputs} are plotted in Figures \ref{fig:blazarEGB}--\ref{fig:zdepcombinedEGB}.  In Figure \ref{fig:blazarEGB}, we have plotted the best-fit spectral shapes of the collective intensities of unresolved FSRQs ($\chi_{\rm red}^2 = 4.1$) and BL Lacs ($\chi_{\rm red}^2 = 3.3$). In Figure \ref{fig:bestcombinedEGB}, we have plotted the best-fit spectral shape of the total collective spectrum of unresolved blazars in Scenario 1 (solid line; $\chi_{\rm red}^2 = 0.7$; for explanation, see Section \ref{sec:inputs}) and the individual contributions from FSRQs (dashed line) and BL Lacs (dot-dashed line). In Figure \ref{fig:zdepcombinedEGB}, we have plotted the best-fit spectral shape of the total collective spectrum of unresolved blazars in Scenario 2 ($\chi_{\rm red}^2 = 0.6$). For comparison, we have also plotted the best-fit spectral shape of the total collective spectrum assuming that FSRQ spectra break at $E_b \sim 1$ GeV and BL Lac spectra break at $E_b \sim 5$ GeV (assuming Scenario 1; Figure \ref{fig:altcombinedEGB}). Note that such spectra do not adequately fit the spectra in Figure \ref{fig:blazarspectra}.  As such, this alternative is included merely for the sake of comparison.

As noted in Section \ref{sec:inputs}, BL Lacs tend to be harder than FSRQs, and thus, the BL Lac collective spectrum is noticeably harder than that of the FSRQs.  Also apparent is the effect of the higher break energy for the BL Lacs as compared with the FSRQs.  As can be seen in Figure \ref{fig:blazarEGB}, the collective spectrum of neither sub-population of blazars reproduces well the spectrum of the EGB:  the BL Lacs are too hard while the FSRQs are too soft.  However, if the spectra are added together as in Figures \ref{fig:bestcombinedEGB} and \ref{fig:zdepcombinedEGB} (renormalized as described in Section \ref{sec:inputs}), the resulting spectrum (solid line) reproduces well that of the EGB (being within the statistical error bars of nearly all of the data points and within $\sim 1.5 \sigma$ of the last data point). The harder spectra and higher break energies for the BL Lacs compensate for the suppressed intensity of FSRQs at higher energies. Notably, the redshift dependence of Scenario 2 shifts the breaks to lower energies resulting in the transition from FSRQ dominance to BL Lac dominance occurring at lower energies ($\sim 1$ GeV rather than $\sim 3$ GeV). Thus, the relative contribution from BL Lacs to the total blazar collective spectrum is greater in Scenario 2 than in Scenario 1.  In the case of the even lower break energies as shown in Figure \ref{fig:altcombinedEGB}, the total collective spectrum fits the EGB fairly well ($\chi_{\rm red}^2= 1.1$), but the relative contributions of the FSRQs and BL Lacs are such that the contribution from BL Lacs dominates at all energies.  This is in contrast with the scenarios presented in Figures \ref{fig:bestcombinedEGB} and \ref{fig:zdepcombinedEGB} in which BL Lacs dominate only energies greater than $\sim$ few GeV. Also, the sharper contrast between Figures \ref{fig:bestcombinedEGB} and \ref{fig:altcombinedEGB} than that between Figures \ref{fig:bestcombinedEGB} and \ref{fig:zdepcombinedEGB} indicate that the transition and the resulting relative contributions of FSRQs and BL Lacs are more sensitive to the break energies than model of the blazar GLF(s). This is due to the fact that each contribution is dominated by its closest and brightest (though still unresolved) members for which the \emph{observed} break energies are similar to their \emph{intrinsic} values for the GLF(s) considered.

\section{Discussion and Conclusion}

We have calculated the spectral shapes of the contributions to the EGB arising from BL Lacs and FSRQs assuming blazar spectra can be described as broken power laws.  We found that in the case that blazar spectral breaks are indeed common neither sub-population alone can adequately reproduce the spectrum of the EGB. However, in a combined spectrum, the harder spectra of the BL Lacs could compensate for the softer spectra of the FSRQs, resulting in a collective blazar spectrum that is similar to that of the EGB within uncertainties. Furthermore, we have found that the relative contributions of FSRQs and BL Lacs required to fit the EGB spectrum are sensitive to the nature of the spectral breaks and, to a lesser extent, the blazar GLFs.

The question of whether the relative contributions necessary to reproduce the overall spectrum of the EGB as determined in this method are reasonable given future {\it Fermi} measurements of the blazar GLFs (which account for the notable observational uncertainties discussed below and in Section \ref{sec:intro}) would provide insight into the overall blazar contribution to the EGB as well as the nature of blazar variability.  According to current {\it Fermi} observations, the luminosities of FSRQs are greater than those of BL Lacs \citep[$L_{\gamma,{\rm FSRQ}} \sim 10^2 \times L_{\gamma,{\rm BLL}}$; ][]{lat09a,agnlat10}.  On the other hand, BL Lacs are likely situated at lower, on average, redshifts than FSRQs \citep{der07}.  Thus, it is possible that the closer proximity of BL Lacs (or their numbers) compensates for their deficit in luminosity with respect to FSRQs resulting in their roughly comparable or slightly enhanced relative contribution to the collective blazar spectrum seen in Figures \ref{fig:bestcombinedEGB}, \ref{fig:zdepcombinedEGB}, and \ref{fig:altcombinedEGB}. However, it should be noted that the closer proximity of BL Lacs and their harder spectral indices would also make them more easily observable by {\it Fermi}, limiting the contribution of \emph{unresolved} BL Lacs to the EGB--though if the luminosity function for BL Lacs is broad enough, there could be many low-luminosity BL Lacs that would escape detection.

It thus remains to be seen how the collective intensities of the sub-populations of blazars actually compare with one another.  In order for a true comparison to be drawn, the GLF for each sub-population of blazars needs to be measured.  If it should be the case that the relative contributions are not reasonable given gamma-ray observations, then either the breaks included in the best-fit spectrum are not typical of blazars or blazars are not sufficient to explain the EGB.  Already, {\it Fermi} studies of observed blazar flux distributions have suggested that emission from unresolved blazars may not comprise the dominant contribution to the EGB \citep{sclat10}. However, given that the physics behind the gamma-ray emission of other known and speculated contributors (i.e., normal galaxies, cascades of ultra-high energy cosmic rays and TeV gamma rays, and dark matter annihilation) result in spectra that are quite distinct from that of the EGB as measured by {\it Fermi} \citep[see e.g.,][]{ando07,kal09,sie09,ahl10,fie10,ven10,ber10,ste10}, a close resemblance of the collective spectrum of unresolved blazars could be striking.

The necessary reconciliation of the clues to the blazar contribution to the EGB provided by studies of the collective blazar spectrum with the clues provided by studies of the blazar source counts could thus provide insight in the gamma-ray emission properties of blazars. Blazar variability plays a substantial role as flaring blazars would be more easily observed by {\it Fermi} than quiescent blazars.  As such, studies of the blazar GLF, observed blazar flux distributions, and ultimately, the blazar contribution to the EGB are largely a question of the blazar duty cycle.  If blazars spend the majority of their lifetimes in the more observationally challenging quiescent state, then studies of the blazar contribution to the EGB based on observed flux distributions such as that presented in \citet{sclat10} could \emph{underestimate} the number of quiescent blazars resulting in an \emph{underestimation} of the blazar contribution to the EGB. Thus, the \emph{apparent} discrepancy of the predictions of the two analyses could be the result of the variability of gamma-ray emission in blazars, the study of which could have implications for gamma-ray emission in blazars.

The comparison of the two types of analyses could also provide insight in the cosmological properties of blazars. At $100$ MeV, the \emph{Fermi}-LAT angular resolution is $\sim 5^{\circ}$ \citep{atw09}; hence, for blazar GLFs that predict large number densities, many blazars, particularly those on the faint-end of the source count distributions, that are, in principle, \emph{observable} would not be distinguishable from other blazars (and thus are \emph{unresolved}) and would not be included in source count distributions. In such a scenario, the faint-end of the blazar flux distribution would underestimate the number of blazars in a given flux bin, and the faint-end slope might appear flatter than it should be\footnote{Another \emph{symptom} (but not proof) of the effect of source confusion would be a similar {\it Fermi} measurement of the EGB as that of EGRET at lower energies since at these energies, the angular resolution of \emph{Fermi}-LAT is comparable to that of EGRET \citep{tho93,atw09,ste10}. Intriguingly, the first few data points of the {\it Fermi} EGB do appear to be consistent with the \citet{str04} determination of the EGRET EGB.}, resulting in an underestimation of the blazar contribution to the EGB. Thus, the \emph{apparent} discrepancy of the results of the two analyses could be the result of a large number density of blazars.

In conclusion, we have demonstrated that even with the inclusion of spectral breaks, the collective spectrum of unresolved blazars reproduces well the spectrum of the \emph{Fermi} EGB for several break models. As such, we find that the possibility that the collective intensity of unresolved blazars dominates the EGB is not excluded on spectral grounds, even if spectral breaks are shown to be common among \emph{Fermi} blazars.  Given the remaining controversy concerning the blazar GLFs, we conclude that it is, as yet, premature to rule out blazar dominance of the EGB. Furthermore, we have shown that relative contributions of the sub-populations of blazars required to fit the EGB spectrum are sensitive to the nature of their breaks; hence, the methodology we present in this paper can be used to constrain the GLFs of blazars. As models for the GLFs of blazars (accounting for the uncertainties outlined in this paper) and more data on spectral breaks become available, the study of the spectral shape of the blazar contribution to the EGB in light of the breaks can provide insight into the high-energy physics of blazars and their cosmological properties.

\acknowledgements{We gratefully acknowledge enlightening discussions with Floyd Stecker.  T.M.V. was supported by an appointment to the NASA Postdoctoral Program at the Goddard Space Flight Center, administered by Oak Ridge Associated Universities through a contract with NASA.  VP acknowledges support for this work provided by NASA through Einstein Postdoctoral Fellowship grant number PF8-90060 awarded by the Chandra X-ray Center, which is operated by the Smithsonian Astrophysical Observatory for NASA under contract NAS8-03060.  This work was partially supported by NASA through the Fermi GI Program grant number NNX09AT74G.}

\clearpage

\bibliography{ms_emulateapj_VP_bibtex}
\bibliographystyle{apj}

\end{document}